\documentclass{article}              
\usepackage{graphicx}
\usepackage{amsmath}
\usepackage{mathtools}
\usepackage{authblk}

\DeclarePairedDelimiter\ket{\lvert}{\rangle}
\DeclarePairedDelimiterX\braket[2]{\langle}{\rangle}{#1 \delimsize\vert #2}

\begin{document}                  


\title{Towards generalized data reduction on a chopper-based time-of-flight neutron reflectometer}


\author[1]{Philipp Gutfreund}
\author[1]{Thomas Saerbeck}
\author[1]{Miguel A. Gonzalez}
\author[1]{Eric Pellegrini}
\author[2]{Mark Laver}
\author[1]{Charles Dewhurst}
\author[1]{Robert Cubitt}

\affil[1]{Institut Laue-Langevin, 38000 Grenoble, France}
\affil[2]{School of Metallurgy and Materials, University of Birmingham, Birmingham B15 2TT, United Kingdom}








\maketitle                        


\begin{abstract}
The calculation of neutron reflectivity from raw time-of-flight data including instrumental corrections as well as improved resolution calculation is presented. The theoretical calculations are compared to experimental data measured on the vertical sample plane reflectometer D17 and the horizontal sample plane reflectometer FIGARO at the Institut Laue-Langevin, Grenoble, France (ILL). This article comprises the mathematical body of the time-of-flight reflectivity data reduction software COSMOS which is used on D17 and FIGARO.
\end{abstract}


\section{Introduction}
\label{sec:intro}
\begin{figure}
\begin{center}
  \includegraphics[width=0.8\linewidth]{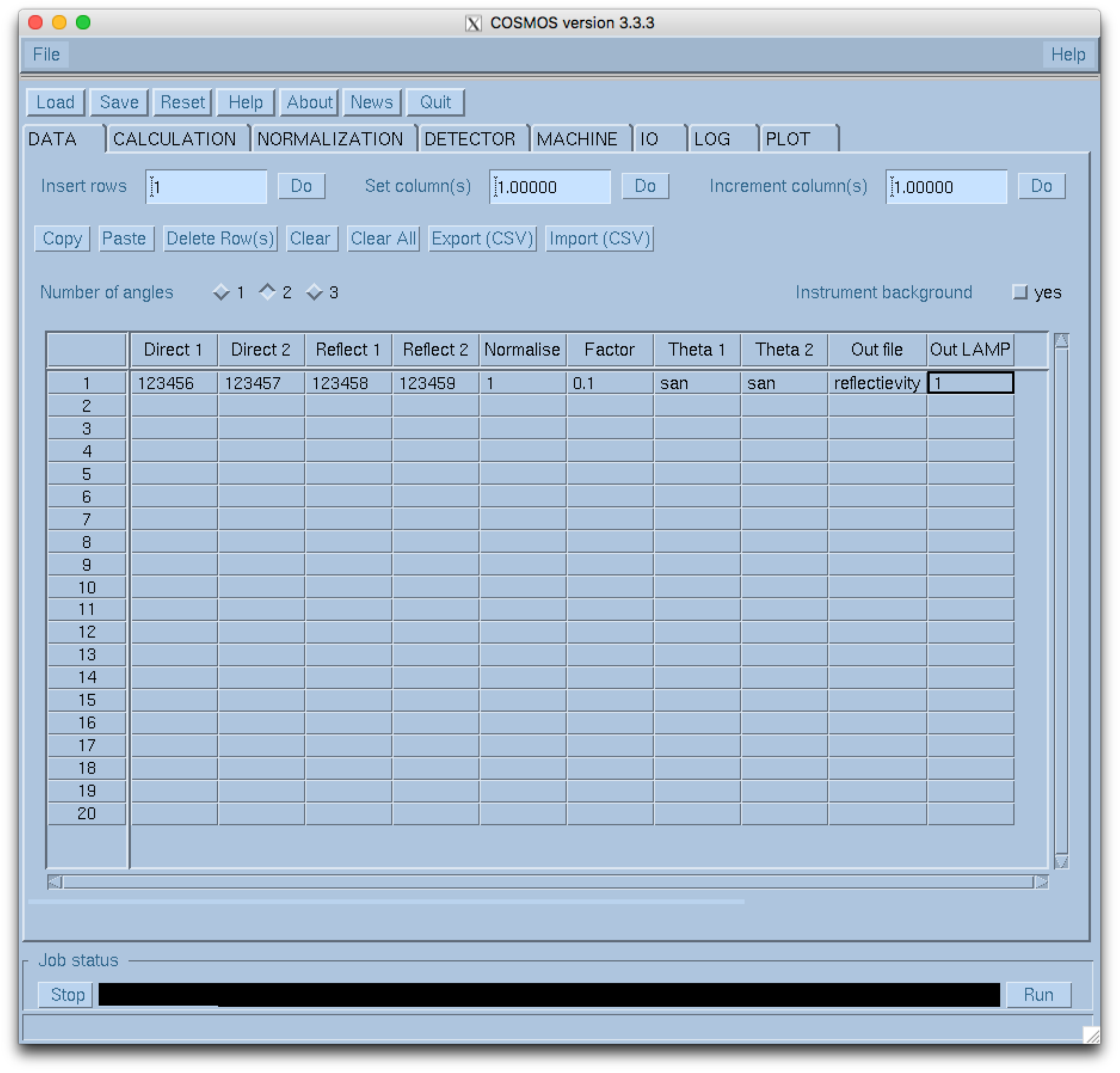}
\caption{Screen-shot of the COSMOS GUI.}
\label{fig:Snapshot}       
\end{center}
\end{figure}

The time-of-flight (ToF) technique is one of the easiest ways to determine the energy and wavelength of neutrons, measuring their speed by timing the neutron flight path. ToF methods were initially used for inelastic neutron scattering at reactor neutron sources. However, the pulsed nature of spallation sources has also brought elastic neutron scattering experiments into play, like ToF - Small Angle Neutron Scattering (SANS) and ToF - Neutron Reflectometry (NR). For continuous sources, a key advantage of using the ToF technique in elastic neutron reflectometry is that a constant fractional momentum transfer resolution can be achieved by using a double chopper system \cite{vanWell1992} where the length of a neutron pulse, and thus the wavelength resolution $\Delta\lambda$, is made proportional to the wavelength. For a constant footprint the beam brilliance at any point on the reflectivity curve can then be maximized by matching the fractional angular resolution $\frac{\Delta\theta}{\theta}$ with the fractional wavelength resolution $\frac{\Delta\lambda}{\lambda}$. On a spallation source however, the lowest wavelength resolution is fixed by the spallation pulse length and/or frequency and by the distance from the source to the detector so there is little flexibility to adapt the wavelength resolution to the experimental problem and/or the angular beam divergence.\\ 
Due to the fact that ToF neutron reflectometry on a reactor source is quite common today, a unified data reduction process is desirable which is already achieved within the ILL by the common ToF-NR data reduction software COSMOS. The program is written in IDL and is called from ILL's Large Array Manipulation Program (LAMP) \cite{LAMP} and communicates via a Graphical User Interface (GUI) for maximal user-friendliness. A snapshot of the GUI is shown in Fig.\,\ref{fig:Snapshot}, where the main tab is seen with an expendable table comprising the experimental run numbers of the direct and reflected beam measurements, which are normalized, merged and exported to a reflectivity curve in ascii format. The other tabs available below the menu include the foreground and background widths and the wavelength range settings as well as the binning factors, automatic normalization calculations, detector masks, instrument parameters, input/output directories, a log and a plot of the final reflectivities. A manual of the GUI can be found here:\,\cite{COSMOSwww}. The present article explains the mathematical body of this data reduction program taking the D17 \cite{Cubitt2002,Saerbeck2018} and FIGARO \cite{Campbell2011} reflectometers as examples.

\section{ToF by using a double chopper}
\label{sec:TOF}
\subsection{Transmitted intensity}
\label{sec:chopperIntensity}
\begin{figure}
\begin{center}
  \includegraphics[width=0.8\linewidth]{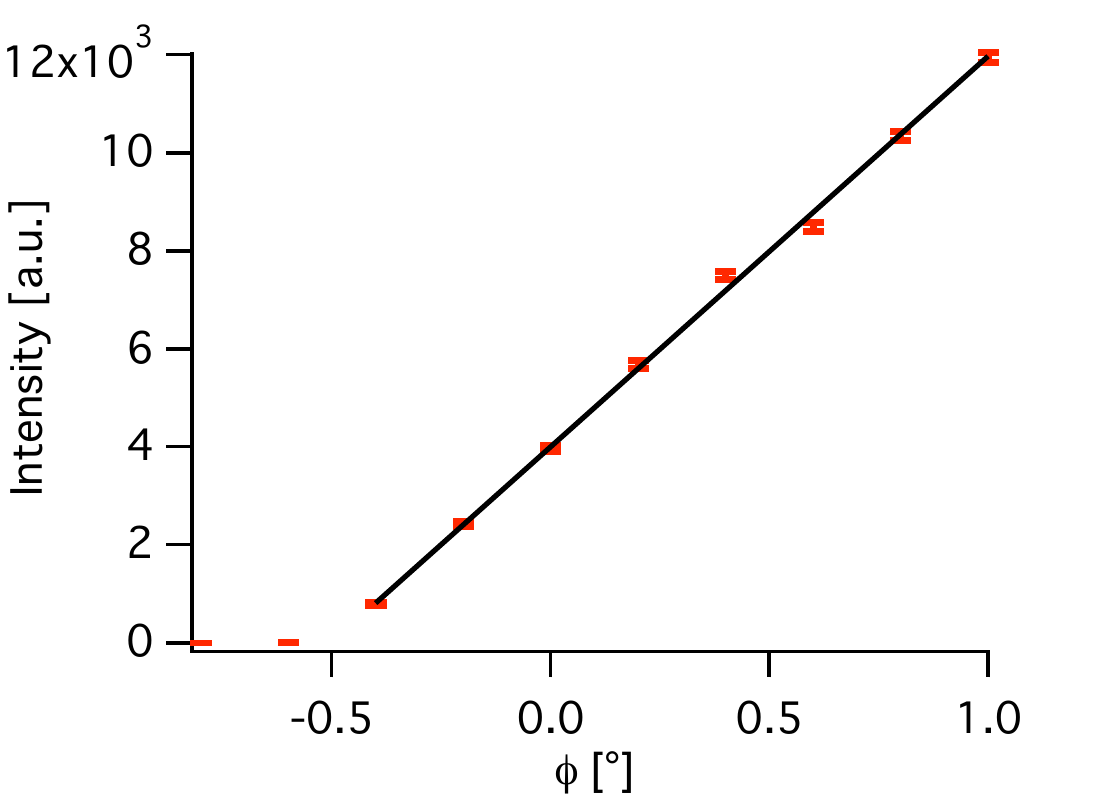}
\caption{Intensity of a monochromatic beam (red bars) as a function of chopper opening measured on D17. The black line is a linear fit.}
\label{fig:OpeningScan}       
\end{center}
\end{figure}

\begin{figure}
\begin{center}
  \includegraphics[width=0.8\linewidth]{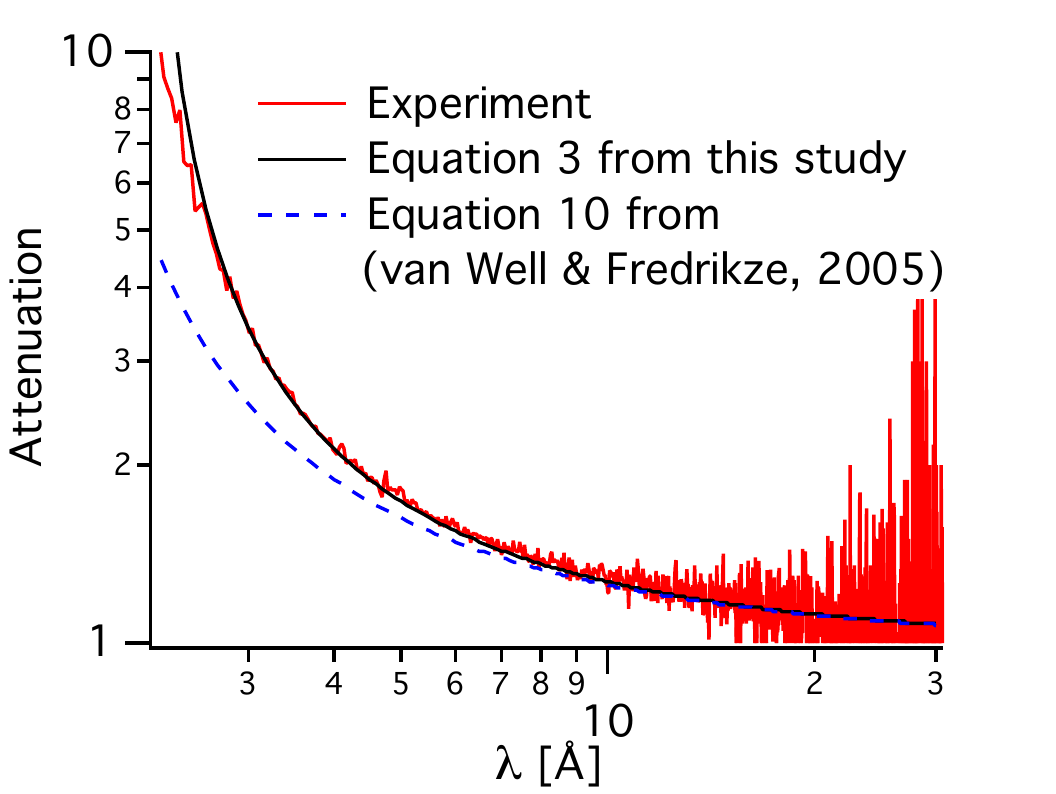}
\caption{The red curve shows the measured transmission on D17 of a white beam at zero opening angle $\phi$ divided by the transmission at a chopper opening of -0.26$^{\circ}$ versus wavelength. The black solid curve is the theoretical result using eq.\,\ref{eq:ChopperTransmission} and the broken line is obtained by using eq.\,10 from \cite{vanWell2005}. Both axes are on log$_{10}$ scale.}
\label{fig:lambdaScan}       
\end{center}
\end{figure}

The functional design of a double chopper system for neutron reflectometry was outlined nearly 30 years ago \cite{Copley1990,vanWell1992} and the corresponding transmission and instrument resolution functions have since been calculated several times \cite{deHaan1995,Cubitt2002,vanWell2005,Campbell2011,Radulescu2015,Pleshanov2017}. However, these calculations have not entirely been corroborated by experimental results and we believe that some additional corrections have to be taken into account in order to match an experimental situation. In this article we only consider set-ups with a single slot chopper system. Multi-slot choppers can effectively decrease the chopper period and thus increase flux, but typical single slot double chopper speeds on a cold neutron source are on the order of 1000\,rpm and thus technically not very demanding. Moreover, a multi-slot double chopper system would have very high requirements on the match up of all chopper slots.\\
The chopper transmission $t$ can be calculated by either computing the wavelength dependent effective opening of the chopper pair and dividing by 360$^{\circ}$ or by calculating the pulse length and dividing by the chopper period $T$. Both calculations lead to the same result:
\begin{equation}
\centering
t=\frac{\Phi_{0}}{2\pi}-\left|\frac{\Phi_{0}-\phi}{2\pi}-\frac{z_{0}\lambda m_{n}}{Th}\right|.
\end{equation}
$\Phi_{0}$ is the transparent sector of the choppers at the beam position, $\phi$ is the opening between the choppers so that a value of zero means that there is no direct line of sight between the choppers and $z_{0},\lambda,m_{n}$ and $h$ are the distance between the two choppers, the neutron wavelength and mass and Planck's constant, respectively. This results in a triangular transmission function with a peak intensity at 
\begin{equation}
\lambda_{0}=\frac{(\Phi_{0}-\phi)hT}{2\pi z_{0}m_{n}}.
\end{equation}
For large sectors of $\Phi_{0}=45^{\circ}$ as on D17 and FIGARO, the maximum chopper separation on FIGARO of 0.8\,m and the lowest chopper speed used of 756\,rpm this value is $\lambda=38$\,\AA\, even for an unusually large opening of 10$^{\circ}$. This is much longer than the maximum wavelength used on the instruments and thus one can safely ignore the influence of $\Phi_{0}$ by going to the short wavelength and small opening approximation which leads to the well-known linear increase of the chopper transmission with neutron wavelength:
\begin{equation}
t=\frac{\phi}{2\pi}+\frac{z_{0}\lambda m_{n}}{Th}.
\label{eq:ChopperTransmission}
\end{equation}
For a fixed wavelength, speed and chopper separation the transmission increases linearly with the chopper opening $\phi$. This is regularly verified on the ILL ToF-reflectometers by measuring the intensity of a monochromatic beam as a function of chopper opening. A typical scan measured on D17 is shown in Fig.~\ref{fig:OpeningScan}.
It can be seen that by over-closing the choppers it is possible to block a certain wavelength completely. This is due to the fact that below a certain threshold wavelength $\lambda_{min}$ the neutrons are too fast to fly through the over-closed choppers. This threshold wavelength can be calculated from equation \ref{eq:ChopperTransmission} for zero transmission:
\begin{equation}
\lambda_{min}=-\frac{\phi Th}{2\pi z_{0}m_{n}}.
\label{eq:OpeningScan} 
\end{equation}
The comparison of the fitted cut-off wavelength in the calibration scan shown in Fig.~\ref{fig:OpeningScan} and the theoretical value is regularly performed in order to calibrate the absolute value of $\phi$ provided that the wavelength is known from a detector distance scan or a chopper speed scan as described in sec.\,\ref{sec:DataRed}.\\
A known result from eq.~\ref{eq:ChopperTransmission} is that the transmission also scales with the chopper speed $2 \pi / T$. However, to avoid the overlap of slow neutrons from one pulse with fast neutrons from the next pulse, the pulse rate cannot be higher than the time needed for the slowest neutrons ($\lambda_{max}$) to travel the chopper-to-detector distance $D_{ToF}$
\begin{equation}
T>\frac{D_{ToF}\lambda_{max}m_{n}}{h}.
\end{equation}
For a typical mid chopper-to-detector distance of $D_{ToF}=7.7$\,m as on D17 and a maximum wavelength of $\lambda_{max}=30$\,\AA\, this leads to a maximum chopper speed of about 1000\,rpm.\\
Instead of increasing chopper speed, another way to gain transmission is to increase the inter-chopper distance $z_{0}$. This leads to a worse resolution, as does increasing the chopper opening. An advantage of opening the choppers is that high resolution is achieved for small momentum transfers, defined as $Q_{z}=\frac{4\pi}{\lambda}\sin(\theta)$, with $\theta$ the reflection angle, while the resolution becomes worse towards the tail of the reflectivity. This may be advantageous as sharp features are found at low $Q_{z}$ values like the total reflection edge and pronounced Kiessig-oscillations whereas at larger $Q_{z}$ the fringes are usually smeared out due to background and roughness. If high resolution is not needed at low $Q_{z}$, a larger inter-chopper distance gives a considerably higher transmission for the same lower end wavelength resolution.\\
The wavelength dependence of the transmission from eq.\,\ref{eq:ChopperTransmission} is shown in fig\,\ref{fig:lambdaScan} and compared with the wavelength dependent transmissions for two chopper openings that were measured on D17 and divided by each other.  
Note that the beam size has no influence on the chopper transmission, in opposition to what has been assumed earlier \cite{vanWell2005}, the wavelength resolution, on the other hand, may be influenced by the beam size as will be shown later. Another experimental validation of eq.~\ref{eq:ChopperTransmission}, especially the invariance to the beam width, was performed by a direct measurement of the chopper transmission on D17 at a fixed wavelength of 5.5\,\AA, where different chopper openings and beam sizes were used (not shown).

\subsection{Wavelength resolution}
\label{sec:lambdaRes}
\begin{figure}
\begin{center}
  \includegraphics[width=\linewidth]{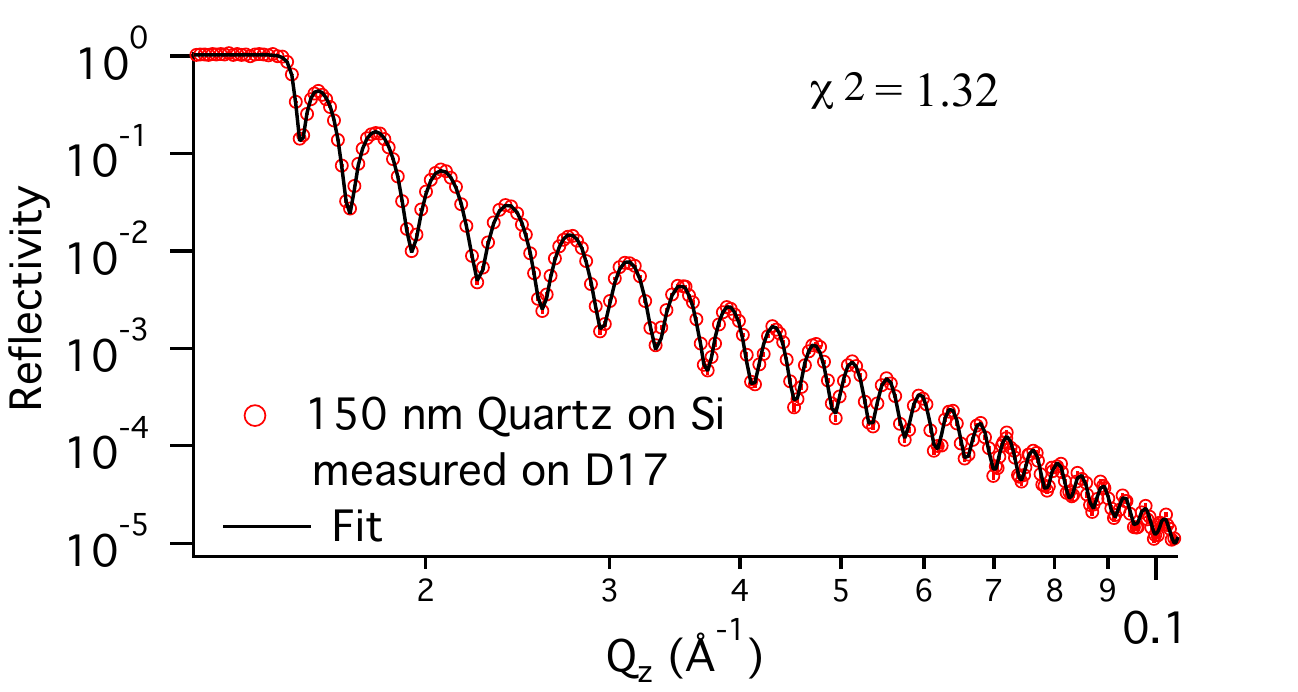}
\caption{NR versus $Q_{z}$ (both on log$_{10}$ scale) from a 150\,nm thick quartz film on silicon, measured on D17 using three angles:  0.8$^{\circ}$, 3.2$^{\circ}$ and 6.2$^{\circ}$ using a wavelength range from 2.5 -- 25\,\AA. The angular resolution was 0.8\% (FWHM) and the wavelength resolution varied from 0.9 -- 1.4\%. The beam footprint was $3 \times 5\,cm^{2}$. Note that only the low $Q_{z}$ part of the reflectivity curve is shown, whereas the fit covers the whole momentum transfer range.}
\label{fig:Quartz}       
\end{center}
\end{figure}

In general, the fractional wavelength resolution is determined by the pulse length defined by the chopper system $\tau_{c}$, the time the choppers need to cut through a beam of size $w$ perpendicular to the chopper movement, the average time the neutron travels through the active zone of the detector and the time bin of the detector electronics, all divided by the ToF of the respective neutron. Moreover a possible variation of the chopper opening can smear the wavelength resolution as well. Usually, all of those contributions are added quadratically \cite{vanWell2005}. This is, however, only correct if all of the contributing resolution functions are Gaussian. In reality none of these contributions are Gaussian: the chopper pulses and the detector binning are both top-hat functions and the beam divergence is usually trapezoidal. This is a general problem in ToF neutron scattering and can be solved by using the exact resolution function in the data analysis. This is, however, computationally intense \cite{Nelson2013} and most available reflectometry analysis programs do not offer this possibility \cite{Motofit}. Therefore Gaussian equivalent widths have to be found for the experimental resolution functions. 
We note that for spallation sources with long pulses the Gaussian equivalent FWHM is not sufficient to describe the wavelength resolution due to the highly non-symmetric pulses in time. In this case the exact resolution function must be taken into account. Accordingly COSMOS saves, on demand, all the relevant instrument parameters needed to calculate the exact resolution function in the header of the reduced data file.\\
The best approximation to experimentally realised smearing is to compare the width of an arbitrary resolution function $R(x)$ (assuming it is symmetric around 0) to an equivalent Gaussian function with the same mean {\it absolute} deviation $<\Delta>$:
\begin{equation}
<\Delta>=\frac{\int_{0}^{\infty}R(x)x\,\textrm{d}x}{\int_{0}^{\infty}R(x)\,\textrm{d}x}.
\end{equation}
This can now be compared to e.g. the Full Width at Half Maximum (FWHM) of a Gaussian function, which is:
\begin{equation}
\textrm{FWHM}=2.9435<\Delta>.
\end{equation}
The resulting FWHM for a top-hat function is 0.736 times the width of the distribution and a trapezoid with a base width of $b$ and a top width of $a$ results in:
\begin{equation}
\textrm{FWHM}=0.49*\frac{b^{3}-a^{3}}{b^{2}-a^{2}}.
\label{eq:TrapezoidFWHM}
\end{equation}
Another possibility to compare the widths of a real resolution function with a Gaussian one is to match the corresponding standard deviations. This leads to a Gaussian equivalent FWHM of 0.69 times the base width of a top-hat function and is usually used to define the resolution on ToF reflectometers \cite{vanWell2005,James2011}. However, measurements on D17 of a highly homogeneous crystal quartz film deposited on a flat silicon wafer (cf. Fig.~\ref{fig:Quartz}) show that the Gaussian equivalent widths calculated by the mean absolute deviation are closer to the real resolution than the ones computed out of the standard deviations. The quality of the fit worsened from $\chi^{2}$ = 1.32 to 1.41 when using the values derived from the standard deviation in the example presented in Fig.\,\ref{fig:Quartz}, while leaving all parameters free to fit. Whilst the difference between using the Gaussian equivalent widths derived by computing the standard deviation and the absolute mean deviation seems to be negligible, COSMOS uses the latter approach to calculate the Gaussian equivalent widths.\\
Depending on the instrument parameters and wavelength the contributions to the wavelength resolution can vary a lot and some contributions may be neglected. The chopper-dependent part of the wavelength resolution is mainly determined by $\tau_{c}$ for cold neutrons:
\begin{equation}
\frac{\tau_{c}}{t_{ToF}}=\frac{1}{D_{ToF}}\left(\frac{\phi T h}{2 \pi\lambda m_{n}}+z_{0}\right),
\end{equation}
with $t_{ToF}$ and $D_{ToF}$ being the time-of-flight and the distance between the middle of the chopper pair and the detection of the neutron, respectively. A chopper opening of $\phi=0$ results in the aforementioned situation of constant fractional wavelength resolution from this term. Therefore it would be advantageous to use a time bin width $\tau_{a}$ which is varied proportionally to the wavelength as well. For convenience, though, a constant time channel width is often used. For 2\,\AA\, neutrons, however, the pulse length on D17 with $\phi=0$ is about 40\,$\mu$s which is even smaller than the typical acquisition channel width of $\tau_{a}=57\,\mu$s, which is chosen to keep the data file size reasonable. Thus $\tau_{a}$ has to be taken into account. As both contributions are top-hat functions the resulting resolution function has the form of a trapezoid and the equivalent Gaussian FWHM can be calculated using eq.~\ref{eq:TrapezoidFWHM}:
\begin{equation}
\frac{\Delta\lambda}{\lambda}=\frac{0.49}{D_{ToF}}\frac{3a^{2}+3ab+b^{2}}{2a+b}, 
\label{eq:lambdares}
\end{equation}
with $a=\frac{\phi T h}{2 \pi\lambda m_{n}}+z_{0}$ and $b=\frac{\tau_{a}h}{\lambda m_{n}}$.\\
Therefore the wavelength resolution is not proportional to the wavelength anymore if a constant time channel width is used. As mentioned earlier this can be improved if the time channel width is varied proportionally to the wavelength. If a fixed number of $N_{ToF}$ time channels are used the corresponding channel width should be:
\begin{equation}
\tau_{a}=\frac{2\lambda m_{n}}{h D_{ToF} N_{ToF}}.
\end{equation}
For one thousand time channels this would lead to a fractional time channel length of 0.2\% of the time-of-flight and thus being negligible in comparison to the wavelength resolution due to the fractional pulse length of about 0.8\% at zero opening.\\
Another possibility for a variable detector time channel width would be to preserve constant $Q_{z}$ steps. This would be particularly interesting for off-specular measurements close to the specular line as this would avoid the distortion of the scattering pattern as it is for ToF reflectometry with constant time channel width. The $Q_{z}$ dependent ToF times in this case are:
\begin{equation}
t_{ToF}(Q_{z}^{n_{ToF}})=\frac{D_{ToF}m_{n}}{h(1/\lambda_{min}-n_{ToF}/N_{ToF}(1/\lambda_{min}-1/\lambda_{max}))}
\end{equation}
for time channel numbers $n_{ToF}$ from 0 to $N_{ToF}$ corresponding to wavelengths $\lambda_{min}$ to $\lambda_{max}$. For $N_{ToF}=1000$ and $D_{ToF}=7$\,m this would correspond to wavelength resolutions of $\Delta\lambda/\lambda=$0.1\% (3.4$\mu$s) and 1.4\% (733$\mu$s) for the limiting wavelengths of 2\,\AA \,and 30\,\AA , respectively.\\
The time the chopper needs to cross a beam of width $w$ at the chopper position is:
\begin{equation}
\tau_{w}=\frac{wT}{2\pi r},
\label{eq:tauw}
\end{equation}
with the chopper radius $r$ at the beam center. The time the choppers need to cut a 1\,cm wide beam at the lowest period used on FIGARO where $T=80$\,ms is about 0.4\,ms for a chopper radius of $r=0.3$\,m and is thus not negligible for large beam sizes. Therefore the FWHM of the chopper crossing time $\tau_{w}$ in eq.\,\ref{eq:tauw} is calculated by estimating the beam cross-section $w$ at the chopper center defined by the two collimating slits from eq.\,\ref{eq:TrapezoidFWHM}. This smearing is subsequently added quadratically to the wavelength resolution from eq.\,\ref{eq:lambdares}.\\
The time a neutron needs to be detected $\tau_{d}$ can be calculated from the width of the active zone in the detector and the absorption length for the given wavelength. As the absorption length inversely scales with wavelength the largest contribution is expected for fast neutrons. The $^{3}$He tube diameter of the D17 and FIGARO detectors is 6.5\,mm. This corresponds to a maximum detection time of 3.3\,$\mu$s. This is much smaller than the usual time channel width and can be therefore neglected.\\
The last influence on the wavelength resolution discussed here is the variation of the chopper opening $\phi$ with time during the measurement. On D17 it is typically less than 0.1$^{\circ}$ (FWHM). This would lead to a change of the chopper pulse of 17\,$\mu$s and is thus much smaller than the 40\,$\mu$s pulse length at 2\,\AA. This would only influence the resolution for chopper settings with an overclosing of more than 0.2$^{\circ}$ which is unusual and is therefore not implemented in COSMOS. 

\section{Data reduction on a ToF reflectometer}
\label{sec:DataRed}
\begin{figure}
\begin{center}
  \includegraphics[width=0.8\linewidth]{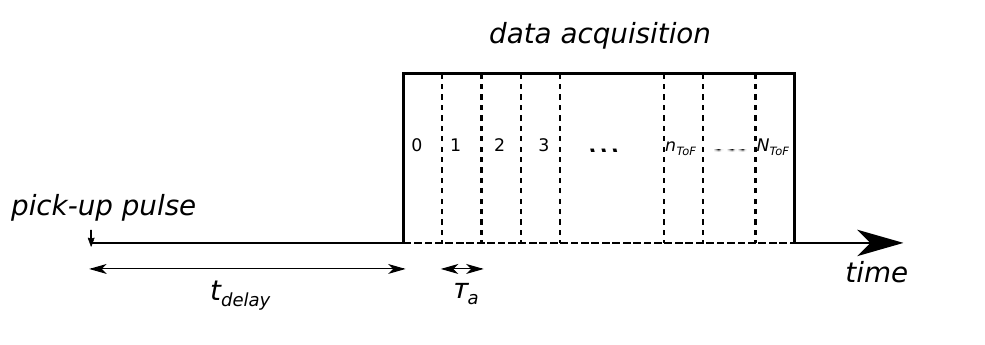}
\caption{Sketch of the detector acquisition schedule as assumed by COSMOS.}
\label{fig:AcquisitionSequence}       
\end{center}
\end{figure}

\begin{figure}
\begin{center}
  \includegraphics[width=0.8\linewidth]{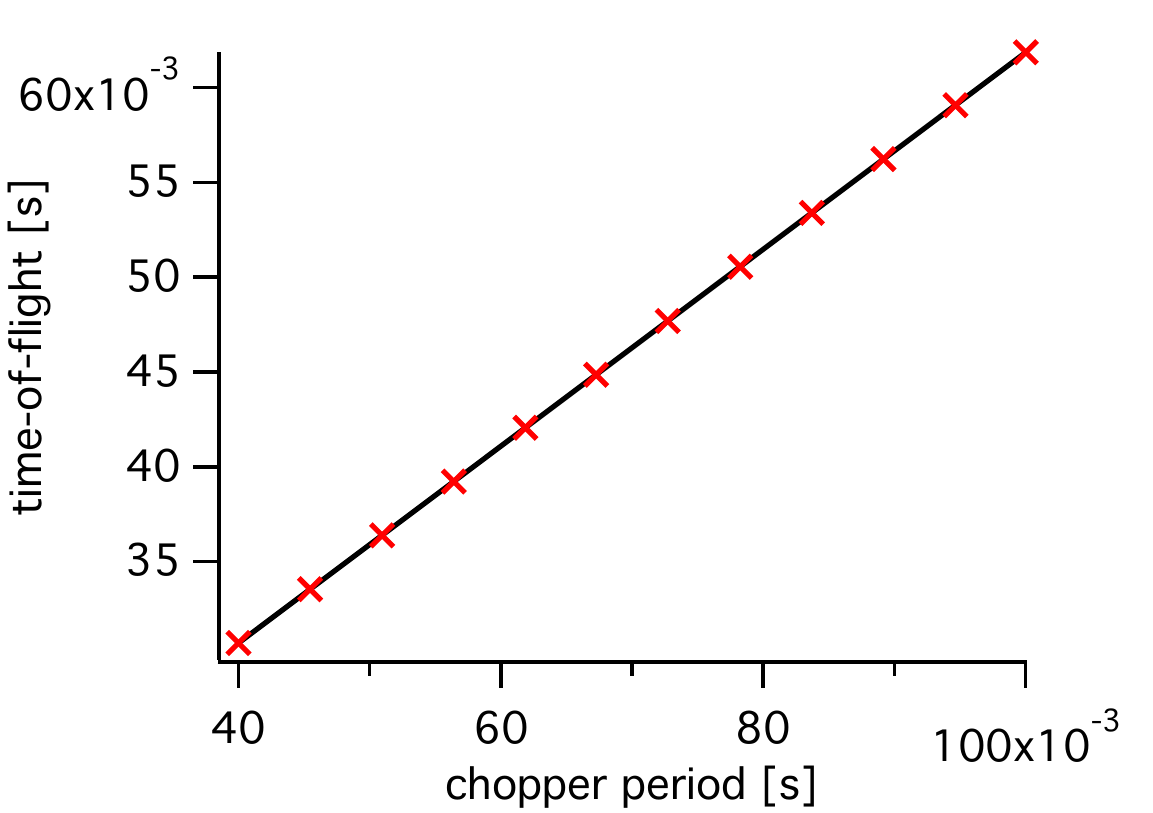}
\caption{Time-of-flight of a monochromatic neutron beam (red crosses) as a function of chopper period measured on FIGARO. The error bars are smaller than the symbols. The black line is a linear fit.}
\label{fig:SpeedScan}       
\end{center}
\end{figure}

In the following the data reduction and possible corrections are explained as they are used for the D17 and FIGARO data reduction software COSMOS. The aim is to produce the normalized specular reflectivity as a function of the normal momentum transfer $Q_{z}$ and to calculate the corresponding statistical errors and momentum transfer resolutions for each point.\\
The wavelength of the detected neutron is calculated measuring the corresponding time-of-flight:
\begin{equation}
\lambda=\frac{ht_{ToF}}{D_{ToF}m_{n}}.
\end{equation}
The ToF distance is calculated by adding the distance from the sample to detector $d_{det}$, the distance from the sample to the leading chopper $d_{0}$ and subtracting half of the inter chopper distance $z_{0}$. All distances are determined by ruler and laser measurements to an accuracy better than 3\,mm. The two chopper discs are equipped with magnetic pick-ups, which send a TTL-type signal at every passage to the detector acquisition system. The pick-up pulse from the first chopper is used to trigger the detector acquisition schedule as sketched in Fig.\,\ref{fig:AcquisitionSequence}. Subsequently the detector acquisition is idle during a certain delay time $t_{dealy}$ which can be set electronically in order to set-up a minimum time-of-flight which corresponds to the shortest wavelength to be recorded. The minimum delay time which comes from signal conversion processes is about 2\,$\mu$s.
If using a constant time channel width $\tau_{a}$ the detector acquisition is sequentially histogramming the detected neutrons into $N_{ToF}$ time channels. The time-of-flight of a neutron registered in the time channel $n_{ToF}$ is calculated according to the following equation:
\begin{equation}
t_{ToF}=\tau_{a}(n_{ToF}+0.5)+t_{delay}-\frac{(\Phi_{off}-(\phi-\phi_{off}))}{4\pi}T
\label{eq:TOF}
\end{equation}
if the first time channel is zero. $\Phi_{off}$ is twice the angle between the trailing edge of the leading chopper blade and the physical pick-up position that sends the electronic start signal to the detector acquisition. It is either calibrated by measuring the time-of-flight of a monochromatic beam of a well known wavelength, determined by a scan of the sample-to-detector distance, to an accuracy of about 0.2$^{\circ}$, as done on D17, or by measuring the time-of-flight of a monochromatic beam (fast enough such that gravity does not play a role) as a function of chopper period as shown in Fig.\,\ref{fig:SpeedScan}, regularly done on FIGARO. According to eq.\,\ref{eq:TOF} the slope of this function is equal to $\frac{(\Phi_{off}^{0}-(\phi-\phi_{off}^{0}))}{4\pi}$. In this way the typical accuracy of determining $\Phi_{off}^{0}$ is 0.05$^{\circ}$.
A possible offset between the pick-ups of the two choppers $\phi_{off}^{0}$ is determined with an opening scan using a monochromatic beam as shown in fig\,\ref{fig:OpeningScan} and compared to the transmission cut-off of the given wavelength from eq.\,\ref{eq:OpeningScan}. The accuracy of this calibration is typically 0.05$^{\circ}$.\\
$D_{ToF}$ and $t_{ToF}$ are both corrected for the flat detector if the neutron arrives at an angle to the normal. The size of the D17 and FIGARO detectors in the out-of-sample-plane direction is 0.25\,m which leads to a maximum correction of 3\,mm at the maximum sample-to-detector distance of 3.1\,m. Another possible correction which is not implemented yet into COSMOS is the wavelength dependent absorption length mentioned in sec.\,\ref{sec:lambdaRes}. At a typical $^{3}$He pressure of 7\,bar neutrons with a wavelength of 27\,\AA\, are detected at 0.5\,mm depth on average, whereas 2\,\AA\, neutrons travel 2.7\,mm through the 6.5\,mm on detection. This would lead to a maximum correction of about 2.2\,mm.\\ 
The reflection angle $\theta_{0}$ can be determined using the sample angle encoder, which relies on the accurate alignment of the sample (typically better than 0.002$^{\circ}$) or by using the detector angle encoder and the position of the reflected beam in comparison to the direct beam to an accuracy of 0.003$^{\circ}$. Both options are available in COSMOS. In total this gives an absolute accuracy of $Q_{z}$ of better than 1\% which is regularly checked on a standard sample. Again, the wavelength dependent absorption length in the detector apparently shifts the position of the beam by an angle of up to 0.002$^{\circ}$ between 2\,\AA\, and 27\,\AA\, neutrons if detected on the very edge of the detector. As the deviation from the wavelength averaged value is only 0.001$^{\circ}$ this correction is negligible on both instruments and not implemented in COSMOS.\\
The resolution in $Q_{z}$ is calculated by summing quadratically the Gaussian equivalent FWHM of the fractional wavelength resolution (see sec.\,\ref{sec:lambdaRes}) and the fractional angular spread of the incoming beam in the simplest case. In this case it is assumed that the sample is underilluminated so that the sample does not act as an additional slit itself. This is reasonable as overillumination should be avoided as it leads to higher background without any gain in reflected intensity. The Gaussian equivalent FWHM of a beam shaped by two collimation slits with sizes $d_{1}$ and $d_{2}$ located at a distance $l$ is given by\cite{vanWell2005}:
\begin{equation}
\Delta\theta^{2}=0.68^{2}\left(\frac{d_{1}^{2}+d_{2}^{2}}{l^{2}}\right).
\end{equation}
The error in determining the Gaussian equivalent FWHM of the resolution introduced by summing squares of non-Gaussian functions as compared to a convolution of the real resolution functions was tested for all possible situations and is below 5\% and thus the use of real divergence resolution functions is not needed for specular reflectometry.
The final fractional $Q_{z}$ resolution is thus:
\begin{equation}
(\Delta Q/Q_{z})^{2}=(\Delta\lambda/\lambda)^{2}+(\Delta\theta/\theta)^{2}.
\label{eq:qres}
\end{equation}
The reflectivity data in ToF mode is collected by using a time-resolved two-dimensional detector. As the neutron beam is usually highly collimated perpendicular to the surface under investigation and divergent parallel to it the scattering pattern is integrated over the parallel direction to reduce the file size. Thus the raw data file reduces to a two dimensional pattern with the projection on the high resolution axis for every time channel.\\   
The reflected intensity as a function of time channel is calculated by normalizing the countrate in a preselected foreground width around the specular peak by the countrate in the same foreground around a direct beam measured with the same conditions as the reflected beam. Optionally a wavelength dependent background can be subtracted from the specular signal by averaging or fitting the countrate in a chosen box around the specular signal for every time channel. If the background becomes $Q_{z}$-dependent as is the case for off-specular scattering for example this procedure is invalid. In this case a constant $Q_{z}$ background reduction has to be applied which will be implemented in the near future. In the more complicated cases, when the sample is not flat or the incoming beam divergence is larger than the detector pixel resolution, COSMOS proposes to use coherent summing, meaning that the foreground is not integrated along lines of constant wavelength as mentioned above but along lines of constant $Q_{z}$. In this case the angular resolution is determined by the smaller of the contributions from the incoming divergence or detector resolution as detailed in Ref.\,\cite{Cubitt2015}.\\
In any case, as the direct beam measurement is done separately, slight differences in slit size, chopper opening and reactor power may influence the normalization. Small influences on the incident neutron flux from the reactor power and feeding guides, which are usually below 5\%, are corrected by using a low efficiency monitor which is placed before the choppers. The actual chopper opening is recorded every 0.3 - 1\,s and the mean value as well as the variance are stored in the raw data files. In case the opening is different for the direct and reflected beam measurements the wavelength dependent chopper transmission is taken into account in COSMOS by using eq.\,\ref{eq:ChopperTransmission}. This correction works very well as shown in fig.\,\ref{fig:lambdaScan}. If different slit settings are used for the direct and reflected beams, COSMOS normalizes the overall counts by the product of the two collimation slit cross-sections. This works quite well for small beam sizes and short wavelengths where the angular beam divergence scales linearly with slit size. For slit sizes larger than 2\,mm or wavelengths longer than 20\,\AA\, this is not true anymore and thus the same slit settings have to be used. If the direct beam becomes too intense for the detector an oscillating slit is used which restricts the height of the beam and acts as an attenuator.

\subsection{Gravity corrections}
In order to account for gravity the raw data for the FIGARO reflectometer are further corrected for the drop of neutrons in the gravitational field. By assuming no change of the final speed of the neutrons due to gravity their trajectory can be described by a parabolic function:
 \begin{equation}
y=y_{0}-k(x-x_{0})^{2},
\label{eq:Parabola}
\end{equation}
where the coordinates $x$ and $y$ describe the horizontal distance towards the neutron source and the vertical height above the center of the sample surface. $k = g/(2v^{2})$ is a characteristic inverse length with the gravitational constant $g$ and the speed of the neutron $v=h/(m_{n}\lambda)$. By imposing two boundary conditions which are the coordinates of two slits before the sample $(x_{1},x_{1}*\tan(\theta_{0})$) and $(x_{2},x_{2}*\tan(\theta_{0}$)) ,with $x_{1}$ and $x_{2}$ being the distance of the two slits from the center of the sample and $\theta_{0}$ the nominal reflection angle at zero wavelength, the two offsets can be calculated:
   \begin{align}
x_{0}=\frac{y_{1}-y_{2}+k(x_{1}^{2}-x_{2}^{2})}{2k(x_{1}-x_{2})} \\
y_{0}=y_{2}+k(x_{2}-x_{0})^{2}.
\end{align}
The position where the neutron hits the sample plane is thus shifted by a factor $x_{s}=x_{0}\pm\sqrt{y_{0}/k}$ where the terms have to be subtracted if the neutron is reflected upwards and added in the case of downwards reflection. The true reflection angle $\theta$ can be hence calculated by differentiating eq.\,\ref{eq:Parabola} with respect to $x$:
   \begin{equation}
\theta=\textrm{atan}(2k*(x_{0}-x_{s})).
\end{equation}
 Finally the chopper pickup offsets have to be re-evaluated:
 \begin{align}
  \Phi_{off}=\Phi_{off}^{0}-(x_{c}*\tan{\theta_{0}} - (y_{0}-k*(x_{c}-x_{0})^{2}))/(2r)\\
  \phi_{off}=\phi_{off}^{0} - \frac{z_{0}}{r}(2k(x_{0}-x_{c})-\tan\theta_{0}),
  \end{align}
  with $r$ being the chopper radius and $x_{c}$ the distance to the middle of the choppers from the sample center. The gravity correction thus leads to a correction of the reflection angle, of the wavelength and directly of the wavelength resolution due to the wavelength dependent opening, all of which is done automatically by COSMOS. 

\subsection{Neutron Polarization Handling}
Neutron beam polarization for experiments on magnetic systems is typically achieved by spin dependent reflection of the neutron beam from a polarizing supermirror. Different designs of supermirrors can be found in the literature, which are all based on the principle of spatial beam separation into $\ket{+}$ and $\ket{-}$ spin states, in which the sign denotes the spin as parallel (+) or antiparallel (-) to the magnetic guide field. Alternative routes for beam polarization or polarization analysis are based on spin dependent absorption in polarized $^3$He \,\cite{Andersen2006, Wolff2006} or refraction in a wedge shaped magnetic field. In a spin polarized neutron reflectometry measurement the detected intensities $I$ can be directly related to spin dependent reflectivities $R$ of the sample by taking into account the polarization setup of the beam. Reflectivities involving only an incoming polarized beam are conventionally described by $R^+$ and $R^-$ for the respective $\ket{+}$ and $\ket{-}$ spin states. Here only the polarizer and first spin flipper are acting on the neutron polarization and only two intensities $I^+$ and $I^-$, are measured. In experiments using full polarization analysis, i.e. the experimental setup includes a polarization analyzer and second spin flipper, the spin state after interaction with the sample is known in addition and separated into non-spin-flip (NSF) $R^{++}$ and $R^{--}$ and spin-flip (SF) $R^{+-}$ and $R^{-+}$ reflectivities \,\cite{Saerbeck2012}. The superscripts denote the spin state before and after the interaction with the sample. D17 operates a polarizing S-Bender \,\cite{Saerbeck2018} in reflection to polarize the beam and a single reflection supermirror or a $^3$He cell for polarization analysis. Two RF spin flippers \,\cite{Grigoriev1997} are available to invert the spin state of the neutron either before or after the sample.\\ The flipping ratio $F=I^+/I^-$ measures the ratio of $\ket{+}$ and $\ket{-}$  states contained in the neutron beam, which is related to the polarization $P$ of a beam with intensity $I_0=I^++I^-$: 
\begin{equation}
P = \frac{I^+-I^-}{I^++I^-}=\frac{F-1}{F+1}.
\label{eq:Pola}
\end{equation}
In investigations of magnetic samples, the sample itself acts as a polarizing element in separating $\ket{+}$ and $\ket{-}$ spin states (NSF reflectivities) or intermixing them (SF reflectivities). For accurate determination of magnetizations and magnetic canting angles the beam polarization has to be taken into account either in the data reduction procedure or during data fitting. The degree of beam polarization provided from a polarizing supermirror depends on the $Q_{z}$ value of the reflection and therefore is angle and wavelength dependent. Monochromatic beam measurements have the advantage of a constant neutron beam polarization if the geometry of the incident beam is not changed during the course of the measurement. A ToF experiment will generally have a wavelength dependent efficiency, leading to a beam polarization varying in $Q_{z}$ with $\lambda$. The procedure for independently determining the wavelength dependent beam polarization has been detailed several times with only small differences in definitions \,\cite{Felici1987, Por1994, Wildes1999, Wildes2006}. By comparing the intensities from two experiments with known spin dependence, the efficiency of spin flippers, polarizer and analyzer can be obtained separately \,\cite{Wildes2006}. Such calibration and control measurements are performed regularly and the results fitted with piecewise linear functions to provide a data independent description of the polarization as a function of wavelength. The piecewise linear function is chosen because of its easy structure and adaptability without having to resort to high-order polynomials.\\
COSMOS provides the option to directly correct recorded intensities for the determined inefficiencies of the devices. The correction uses matrix multiplication of the inverse efficiency matrices and the grouped vector of recorded spin states,
\begin{align}
\hat R &= \hat a^{-1} \cdot \hat p^{-1} \cdot \hat F_2^{-1} \cdot \hat F_1^{-1} \cdot \hat I
\label{eq:polmatrix}
\end{align}
in which $\hat p$, $\hat a$, $\hat f_1$ and $\hat f_2$ represent the spin efficiency matrices from \,\cite{Wildes1999}.
For a full accurate correction, all four intensities $I^{++}$, $I^{--}$, $I^{+-}$ and $I^{-+}$ have to be recorded. However, in most cases the $R^{+-}$ and $R^{-+}$ reflectivities are equal and no new insights in the magnetic order are gained by measuring both cross-sections \,\cite{Zabel2007}. An efficiency correction on a shortened measurement can be performed under the assumption that $R^{+-} \equiv R^{-+}$, which allows one to calculate the expected intensity and the remaining reflectivities from eq.\,\ref{eq:polmatrix}. Equally, missing intensities can be calculated if only the non-spin-flip intensities $I^{++}$ and $I^{--}$ are known, but with the assumption of $R^{+-} \equiv R^{-+}\equiv 0$. This case is rare, as the same information is obtained in a measurement without analysis, i.e. recording $I^+$ and $I^-$. In this case, the efficiency corrections only take into account the polarizer and first spin flipper in a simplified matrix equation. 
\begin{align}\label{spolmatrix}
\left(
\begin{array}{c}I(0)\\I(1)\end{array}\right)&=
\left[\begin{array}{cc}1&0\\
(1-F_1)&F_1
\end{array}\right]\times\left[\begin{array}{cc}
(1-p)&(p)\\
(p)&(1-p)\\
\end{array}\right]
\left(\begin{array}{c}R^{+}\\R^{-}\end{array}\right).
\end{align}
Because the polarizer and analyzer only create a wavelength dependent scaling, it is sufficient to record a direct beam with $I^{00}$ setting to perform the data reduction. All four channels are binned to the same $Q_{z}$-bins by using the same integration range and peak location on the detector. COSMOS applies the appropriate corrections automatically after testing the datafiles for compatibility and detecting how many different spin states are supplied. The data is binned and background subtracted prior to correction in order to provide better statistics. Each correction includes a full error calculation, which is based on the errors determined during efficiency calibration. This procedure typically allows one to measure flipping ratios of 1000, i.e. spin-flip intensities three orders of magnitude lower than the non-spin-flip intensity. Below this, statistical errors in the efficiency evaluation and the instrumental background have too large of an effect to provide physically meaningful data in reasonable measurement times. Here a monochromatic measurement may reach lower values due to the better known efficiency due the peak flux. Uncertainties in the spin-dependent and spin-independent background, however, remain an issue. Measurements of the efficiency with beams of different divergence and beam dimensions showed no effect in the S-Bender and negligible effects from the analyzer supermirror within the typical experimental conditions. 
\begin{figure}
\begin{center}
  \includegraphics[width=0.8\linewidth]{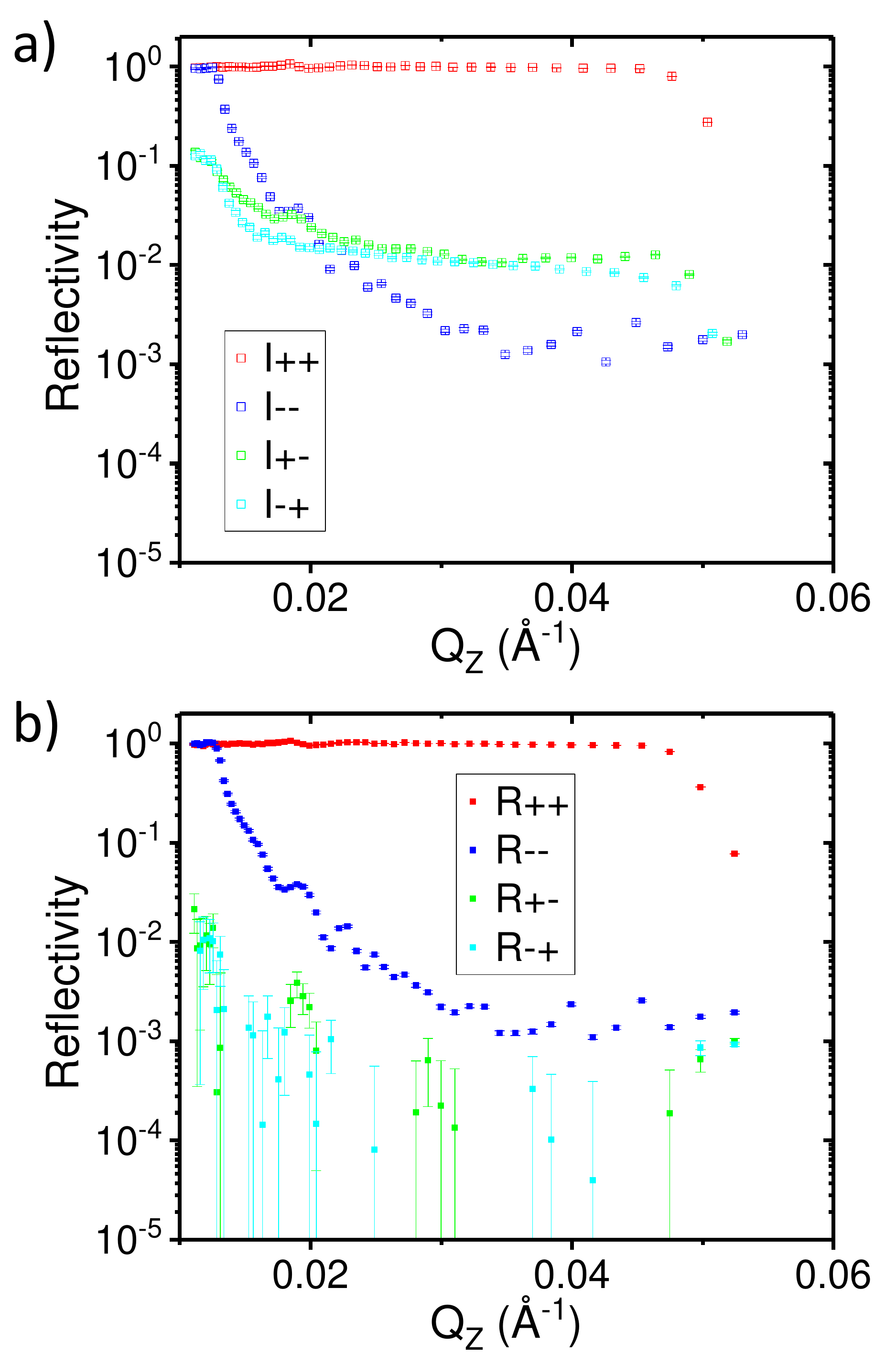}
\caption{Reflectivity measured on D17 from an m = 2.8 Fe/Si supermirror before (a) and after (b) applying the efficiency correction in the data reduction.}
\label{fig:polCOSMOS}       
\end{center}
\end{figure}
Figure \ref{fig:polCOSMOS}a shows the spin resolved intensity reflected of an m=2.8 Fe/Si supermirror saturated in a field of 1 T recorded on D17. This sample acts as an efficient polarizing element itself when inserted into the neutron beam, leading to distinct features observed in the uncorrected intensities. Below the critical edge of total reflection for $\ket{+}$ and $\ket{-}$ neutrons, the comparably low analyzer efficiency for long wavelength neutrons ($a\geq 90\%$) leads to an intensity of $I^{+-}\sim I^{-+}\sim 0.1$, while both $I^{++}$ and $I^{--}$ are normalized to unity. For decreasing wavelength the efficiency of the analyzer improves, but also the reflectivity of $\ket{-}$ neutrons from the supermirror sample decreases rapidly. This means the analyzer is no longer the determining element, as both sample and analyzer predominantly reflect $\ket{+}$ neutrons. Instead, polarizer and spin flipping efficiency of the RF flippers have a larger effect on the intensity distribution. At sufficiently high $Q_{z}$, the spin-flip intensities become larger than the $I^{--}$ intensity. This illustrates that flipping ratios of $FR=I^{++}/I^{--}=1000$ can be measured even though the beam polarization is on the order of 99\% - 98\%, i.e. more than an order of magnitude worse. The difference between $I^{+-}$ and $I^{-+}$ is a result only of the different wavelength dependence of the efficiency of the elements.\\
The result of the data correction using the inverse efficiency matrices from \,\cite{Wildes2006} is shown in Figure \ref{fig:polCOSMOS}b. Only a small effect is observed in the $R^{++}$ and $R^{--}$ channels, which can now be related directly to the polarization efficiency, or magnetization, of the sample. The $R^{+-}$ and $R^{-+}$ channels decreased to the value of the background created by the intensity in the $R^{++}$ and $R^{--}$ channels, whose statistical uncertainty dramatically affects the exact subtraction of spin-polarized contaminations.

\subsection{Data binning}
Due to simplicity usually a constant time channel width is used in the detector acquisition on D17 and FIGARO. This leads to the situation that the time channel width is much shorter than the pulse length for long wavelengths. Those time channels can be binned in order to reduce the number of points with negligible resolution loss. This possibility is available in COSMOS and is implemented in the following way. The algorithm creates the first $Q_{z}$-bin $Q^{bin}_{0}$ and sums up all counts from $Q_{z}$-values between the first unbinned point $Q_{0}$ until $Q_{n}=Q_{0}+\Delta Q_{0} * f$ with $f$ being an input binning factor and $\Delta Q_{0}$ the $Q_{z}$-resolution (see eq.\,\ref{eq:qres}) of the first point $Q_{0}$. The $Q_{z}$-value of the final bin is:
\begin{equation}
Q^{bin}_{j}=\sum_{i=0}^{n}Q_{i}/(n+1).
\end{equation}
As binning is effectively a convolution with a top-hat function the final $Q_{z}$-resolution of the binned point is calculated in the following way:
\begin{equation}
\Delta Q^{bin}_{j}=\sqrt{\sum_{i=0}^{n}\Delta Q_{i}^{2}+((Q_{n}-Q_{0})*0.76)^{2}}.
\end{equation}
This algorithm is then sequentially performed on all data points until the last unbinned point is reached. Care is taken that the statistical counting error calculation is done on the binned data points (if the binning option is chosen) in order to minimnize the influence of zero counts.

\section{Outlook}
Several improvements of the usage of 2D time-of-flight neutron reflectivity patterns are planned in future, and will be incorporated in COSMOS. Most of them relate to the newly developed coherent summing method \cite{Cubitt2015} where the detector resolution is used to partially recover the resolution loss of a highly divergent incoming beam or a bent sample. The first upgrade tackles the issue of wavelength resolution smearing due to the finite beam width at the chopper position as described in Sec.\,\ref{sec:lambdaRes}. As the position-sensitive detector effectively records a pinhole image of the divergent source the neutrons can be tracked back in space and time to the chopper blade position and the smearing can be partially corrected similar to the coherent method. The second upgrade concerns the gravity correction in the coherent method, which at the time of writing is only partially integrated in COSMOS. This will make the use of this method available for reflection down measurements on FIGARO, which are only possible for short wavelengths at the moment. The final improvement of the coherent method involves a point-by-point normalization (and resolution calculation) of the reflected to the direct beams, which will make arbitrary beam profiles accessible. This will become important for the foreseen focusing guide upgrade on D17 \cite{Saerbeck2018}, where the incoming beam divergence will be increased by a factor of three, potentially accompanied by a non-symmetric beam profile. At the moment the coherent option assumes a symmetric beam profile as every pixel in the reflected beam is normalized to a single integrated number of the direct beam flux. The last improvement concerning the coherent method involves generalizing the code to additionally read 3D data files (x vs. y vs. ToF), which would make it possible to handle arbitrarily bent samples; currently COSMOS can only handle samples bent along the reflection plane.\\
Further general improvements of COSMOS include a constant $Q_{z}$ background reduction. We also plan to translate the code from the current IDL programming language to Python, with the aim of integrating the program into Mantid \cite{Mantid}.


\appendix
\section{Calculation of beam footprint for a horizontal sample plane reflectometer}

The footprint of the neutron beam produced by two slits is a trapezoidal intensity distribution along the x-axis of the sample defined by four inclination points: $r_{1},r_{2},l_{1},l_{2}$. The fractional intensity is 0 for $x<l_{1}$, $(x-l_{1})/(l_{2}-l_{1})$ for $l_{1}<x<l_{2}$, 1 for $l_{2}<x<r_{1}$, $(r_{2}-x)/(r_{2}-r_{1})$ for $r_{1}<x<r_{2}$ and 0 for $x>r_{2}$. The fractional illumination in \% is thus given by 100\%*($r_{2}-l_{1})/l_{0}$, where $l_{0}$ is the length of the sample.\\
The wavelength dependent footprint shift $x_{s}$ can be calculated in the following way: 
\begin{equation}
\begin{array}{l}
y_{1}=\tan(\theta_{0})*x_{1}+dy_{2}\\
y_{2}=\tan(\theta_{0})*x_{2}+dy_{1}\\
v=3956/\lambda\\  
k=g/(2*v^{2})\\ 
x_{0}=((y_{1}-y_{2})/k-(x_{2}^{2}-x_{1}^{2}))/(2*(x_{1}-x_{2}))\\ 
y_{0}=y_{1}+k*(x_{1}-x_{0})^{2}\\   
x_{s}=x_{0}\pm\sqrt{y_{0}/k}  
\end{array}
\end{equation}
where $x_{1}$ is the distance from the sample to sample slit in m, $x_{2}$ is the distance from the sample to the collimation slit in m, $\theta_{0}$ is the nominal reflection angle, $\lambda$ the neutron wavelength in \AA\, and the gravity constant $g=9.81$\,kgm/s$^{2}$. The terms for $x_{s}$ have to be subtracted for reflection up and added for reflection down geometry. Finally the trapezoidal inclination points in mm can be calculated:
 \begin{equation}
\begin{array}{l}
l_{1}=x_{s}(dy_{1}=-d_{1}/2000,dy_{2}=d_{2}/2000)*1000+l_{0}/2\\
l_{2}=x_{s}(dy_{1}=d_{1}/2000,dy_{2}=d_{2}/2000)*1000+l_{0}/2\\
r_{1}=x_{s}(dy_{1}=-d_{1}/2000,dy_{2}=-d_{2}/2000)*1000+l_{0}/2\\
r_{2}=x_{s}(dy_{1}=d_{1}/2000,dy_{2}=-d_{2}/2000)*1000+l_{0}/2
\end{array}
\end{equation}
with the slit widths $d_{1}$ and $d_{2}$ in mm.
The Gaussian equivalent FWHM divergence of a beam shaped by two collimation slits with sizes $d_{1}$ and $d_{2}$ located at a distance $l=x_{2}-x_{1}$ is given by\cite{vanWell2005}:
\begin{equation}
\Delta\theta^{2}=0.68^{2}\left(\frac{d_{1}^{2}+d_{2}^{2}}{l^{2}}\right)
\end{equation}
which results in a fractional angular resolution in \% of 100\%*$\Delta\theta/\theta$.



\section*{Acknowledgements}
The authors acknowledge the help of Erik Watkins, Richard Campbell, Robert Barker and Giovanna Fragneto during the improvements and testing of COSMOS. The valuable comments of Armando Maestro are acknowledged. We acknowledge financial support from SINE2020 for the ongoing conversion of COSMOS into Mantid. 


\bibliographystyle{plain}






\end{document}